\documentclass{article}
\pdfoutput=1

\usepackage[preprint,nonatbib]{neurips_2018}

\usepackage{float}
\usepackage[utf8]{inputenc} 
\usepackage[T1]{fontenc}    
\usepackage{hyperref}       
\usepackage{url}            
\usepackage{booktabs}       
\usepackage{makecell}
\usepackage{amsfonts}       
\usepackage{nicefrac}       
\usepackage{microtype}      
\usepackage{amsmath} 
\usepackage[flushleft]{threeparttable}
\usepackage{threeparttable}
\usepackage{bm}
\usepackage{graphicx}

\title{ASBERT: Siamese and Triplet network embedding for open question answering}

\author{%
  Olabanji Shonibare\\
  Kindle on Alexa\\
  \texttt{\{olabanjs\}}@amazon.com \\
}

\begin{document}

\maketitle

\begin{abstract}
Answer selection (AS) is an essential subtask in the field of natural language processing with an objective to identify the most likely answer to a given question from a corpus containing candidate answer sentences. A common approach to address the AS problem is to generate an embedding for each candidate sentence and query. Then, select the sentence whose vector representation is closest to the query's. A key drawback is the low quality of the embeddings, hitherto, based on its performance on AS benchmark datasets. In this work, we present ASBERT, a framework built on the BERT architecture that employs Siamese and Triplet neural networks to learn an encoding function that maps a text to a fixed-size vector in an embedded space. The notion of distance between two points in this space connotes similarity in meaning between two texts. Experimental results on the WikiQA and TrecQA datasets demonstrate that our proposed approach outperforms many state-of-the-art baseline methods.
\end{abstract}

\section{Introduction}

The ability of machines to have a seamless conversation with any human is one of the objectives of Artificial intelligence. Over the years, there has been lots of advancement in the deployment of practical conversational systems such as Amazon's Alexa, Google Assistant, Apple's Siri and Microsoft's Cortana. A common feature included in majority of these system is a Question Answering (QA) unit which attempts to provide a correct answer to any natural question posed to it by the user. In general, question answering can be differentiated in two major ways: Open-domain question answering (QA) and Closed-domain question answering (QA). While closed-domain QA covers questions limited to a particular field such as Law, Investment, e.t.c, open-domain QA has no such restrictions.

An open-domain QA system typically comprises the following pipeline stages:  A user question is parsed to determine its type and/or extract keywords; find suitable documents from a very large corpora; for each selected document, identify candidate answer sentences; lastly, determine a subset of the list that most likely contains the answer to the given question. In this work, our primary focus is on the task of selecting and ranking plausible answers to a given question from a set of candidate sentences, which is often referred to as answer  election. More formally, given a question sentence, $ y $,
and a set containing candidate answer sentences, $ \{x_{i}\}_{i=1}^m $, the objective is to identify the sentence, $ x_j $, that most probably contains the correct answer. A key difficulty associated with this effort is that the most suitable answer sentence might contain a lot of unrelated information and shares very few lexical features with the question sentence. An example is illustrated in Table~\ref{tab:introduction}. One other interesting application of answer selection different from question answering is Information extraction and Knowledge base construction.

Distributed word encodings have been demonstrated to be effective in multiple NLP tasks. However, there are some classes of problems where these word vector representation are inadequate. To handle complex tasks like Machine translation, Question answering and Semantic textual similarity, sentence encodings are desired for a better language comprehension. State-of-the-art models like BERT \cite{bert2018}, RoBERTa \cite{roberta2019} and XLNet \cite{xlnet2019} have shown great performance in several benchmark tasks including question answering, however, it can be computationally expensive when used for text regression problems  since it involves a huge number of sentence comparisons. In addition, the embeddings generated either by averaging the output vectors or using only the vector that corresponds to the first token, [CLS], for the case of a single sentence input, has been shown to be of low quality \cite{reimers2019}.

In this article, we present ASBERT, a deep learning framework based on BERT that utilizes both siamese and triplet networks to learn useful sentence encodings. We demonstrate the superiority of this approach by comparing its performance with strong baselines based on deep learning and cutting-edge sentence encoding functions on the following benchmark datasets for answer selection: TrecQA \cite{wang2007} and WikiQA \cite{yang}.

The remainder of the paper is structured as follows:  Section \ref{related_work} describes related work; Section \ref{approach} presents ASBERT and the different architectures studied; The experimental settings, results and discussions are all presented in Section \ref{experiments};  and finally, we conclude in Section \ref{conclusion}.

\begin{table}[!hb]
	\centering
	\begin{threeparttable}
		\caption{An example of a plausible answer to a query containing lots of redundant information}
		\label{tab:introduction}
		{\small 
		\begin{tabular}{lp{5.6cm}c}
			\toprule
			Query & Sentence    & Label \\
			\midrule
			what causes heart disease & 
			
			Cardiovascular disease is the leading cause of deaths worldwide, though since the 1970s, cardiovascular mortality rates have declined in many high-income countries.
			
			& not relevant     \\
			& There is therefore increased emphasis on preventing atherosclerosis by modifying risk factors, such as healthy eating , exercise , and avoidance of smoking.
			
			& not relevant     \\
			& Cardiovascular disease (also called heart disease) is a class of diseases that involve the heart or blood vessels ( arteries , capillaries , and veins ).The causes of cardiovascular disease are diverse but atherosclerosis and/or hypertension are the most common.     & relevant  \\
			\bottomrule
		\end{tabular}
	    }
		\begin{tablenotes}
			
			\item The label \textit{relevant} and \textit{irrelevant} connotes whether or not the corresponding candidate sentence contains an aswer to the question
		\end{tablenotes}
	\end{threeparttable}
\end{table}

\section{Related work}
\label{related_work}

Over the years, there has been lots of contribution towards automating answer selection from a corpus. Early attempts \cite{heilman2010,wang2015} approached this problem as an approximate matching with n-grams between a query and a candidate answer sentence or applied techniques used to recover the orginal order of a shuffled sentence and they were based on handrafted features. A major drawback with this approach is that it is difficult to generalize, in practice, across different datasets or tasks.


Some other works include the use of a quasi-synchronous dependency grammar to learn a function that maps a query to a candidate answer sentence based on their sentence structure; Heilman and Smith~\cite{heilman2010} suggested an approach to model the similarity of sentence pairs using Tree Edit Distance method (TED); Around the same period, a probabilistic tree-edit model \cite{wang2010} was proposed to determine textual entailments and answers to questions; Furthermore, a linear-chain conditional random field \cite{yao2013} was employed to learn a function that maps a query to a candidate answer sentence employing features obtained from TED.

In recent years, there has been a surge in the adoption of deep neural networks to automate answer selection. Most of these networks can be broadly classified into the following categories: Siamese networks, Attentive networks and Compare-Aggregate networks. A Siamese network consists of two identical subnetworks which are trained to learn the similarity of two input vectors. Only one of these subnetworks is used offline. This network was initially developed for the verification of signatures \cite{bromley1994}.
It was later extended for other use cases including face verification \cite{chopra2005},
reducing the dimension of image feature vectors \cite{chopra2005} and answer selection \cite{conneau2017,reimers2019}. Attention-based frameworks \cite{rockt2015, hermann2015, tan2016} accomodates the influence of other sections in the estimation of the contextual representation of a particular section of an input text. The effectiveness of these systems have been demonstrated across many NLP problems including machine reading comprehension \cite{hermann2015}, text entailment \cite{rockt2015} and question answering \cite{tan2016}. A compare-aggregate system \cite{wang2016, parikh2016} starts, in general, with the splitting of the input text into fewer parts. An encoding of each part is compared and the result of this comparison is aggregated to derive the final distributed representation of the input. The practicality of this framework in solving AS problems has been demonstrated in a number of papers \cite{bian2017,yoon2019,zheng2018}.

Only very few works in the literature have explored the use of a triplet network to address the AS task 
and many of those are based on either an LSTM, CNN or a combination of both structures. To the best of our knowledge, this work appears to be the first to evaluate the embeddings obtained from a siamese/triplet network built on a BERT architecture for AS problems.

\section{Approach}
\label{approach}
\begin{figure}[!hb]
	\centering
	\includegraphics[scale=0.5]{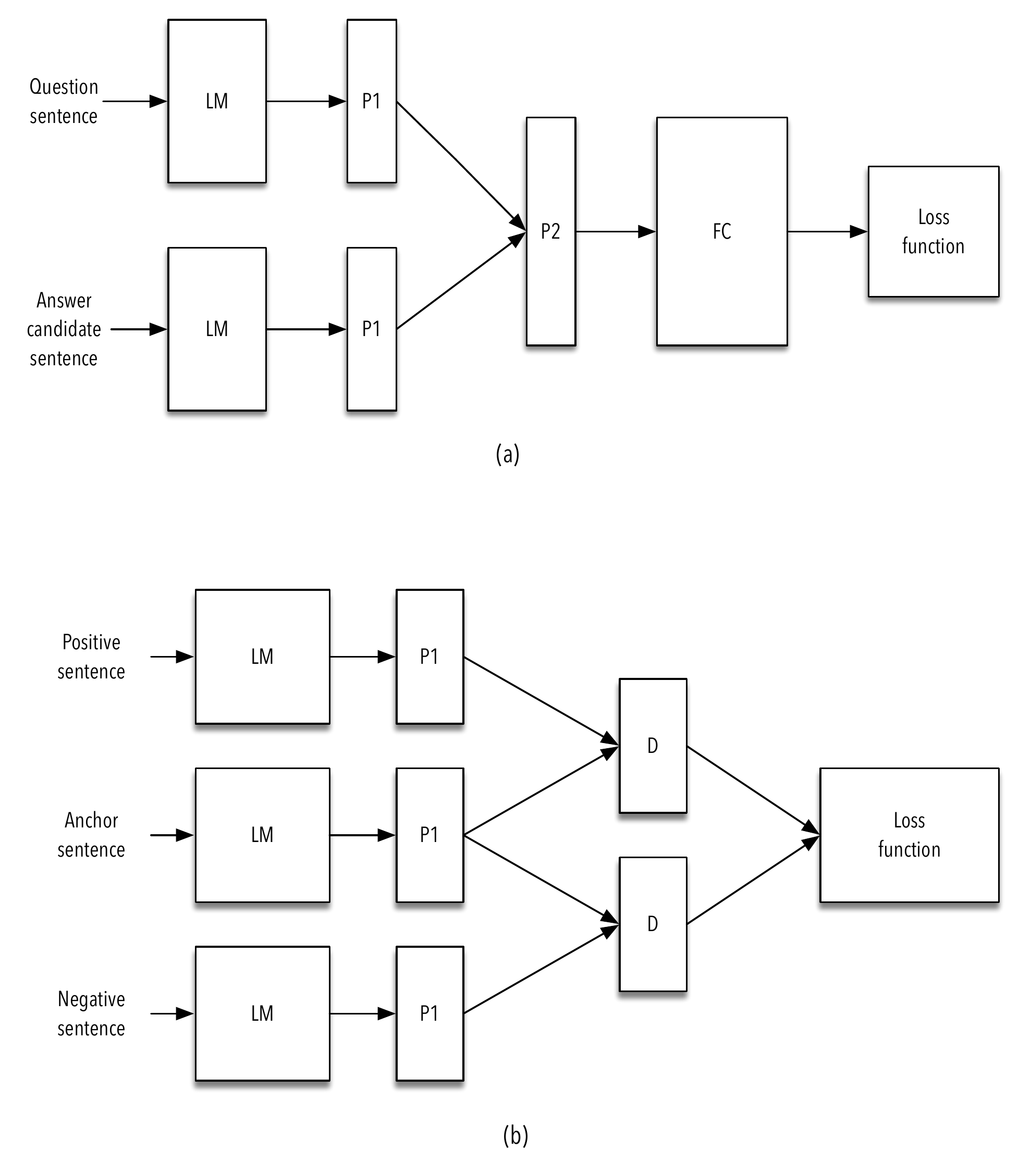}
	\caption{Siamese (a) and Triplet (b) network architecture. LM is a pretrained language model, P1 and P2 are pooling layers, FC is a fully-connected layer and D represents a distance function.}
	\label{fig:architecture}
\end{figure}

We focus on two different architectures: The Siamese and the Triplet networks. The underlying goal of both networks is to learn a semantic embedding space into which an input text could be mapped. This way, given a question sentence and a candidate answer sentence, a distance function can be applied to their vector representation to determine their closeness. The term, closeness, in this work refers to how well the candidate sentence answers the question. The general architecture is shown in Figure~\ref{fig:architecture}.

For both architectures, LM is a pretrained language model which is based on a transformer architecture. The Transformer comprises many attention blocks and within a block, an input vector is transformed by a combination of a self-attention layer and a feed-forward neural network in that order, where the self-attention layer incorporates the influence of neighboring words to the encoding of a particular word. The forward propagation begins with the computation of three embeddings: the token, the position and the segment embeddings. The token embeddings are obtained by splitting the input sentence into tokens and then each token is replaced with their corresponding index in the tokenizer vocabulary; the position embeddings represents the relative postion of each token within a sentence while the segment embeddings is used to address a situtaion where the input is composed of two sentences and is obtained by assigning the tokens for each sentence a unique single label e.g. $ 0 $ and $ 1 $. The final input is then obtained by adding these embeddings. This sequence is mapped through several layers up to the last. The final output, which is the same size as the input embeddings is passed to a pooling layer (P1). 
Some examples of pooling operations that can be applied here includes, max, mean, extracting only the sequence that corresponds to the first token ([CLS]) or even applying attention to the sequence. In this work, the mean pooling operation is adopted. The result of the pooling operation in P1 is then passed on to a number of layers downstream before finally applying a loss function. In Figure~\ref{fig:architecture}, the operations in the two pooling layers, P1 and P2, does not have to be the same. Once the loss optimization procedure is complete, the trained base network LM-P1 would have learnt a function that is able to generate discriminative features, given any sentence, such that the distance between a question and a positive answer sentence is small in the embedding space and large for the same question and negative answer sentence. The details of each network are delineated below.

\subsection{Siamese network}

In this paper, we address the problem of learning an encoding function as a supervised learning task. Each training example comprises a question sentence, a candidate answer sentence and  a label indicating whether it is a positive match $ (1) $ or a negative match $ (0) $. The two LM modules shown in Figure~\ref{fig:architecture} are essentially copies of the same network. They share the same parameters. During training, the LM-P1 module accepts two text inputs, $ x_{1} $ and $ x_{2} $ and produces vectors $ h_1(x_{1}) $ and $ h_1(x_{2}) $, respectively. The output of both network is accepted by the pooling layer, P2, to produce another vector, $ v = h_2(h_1(x_1),h_1(x_2)) $, say. The vector, $ v $, is then transformed by the fully-connected layer to generate the final encoding. If the network is trained optimally, we would expect the distance between generated feature vector for a pair with a positive match to be small and large when it is a negative match.

\paragraph{Loss function}

Given a minibatch containing $ m $ pairs of question sentence $ (x_1) $ and candidate answer sentence $ (x_2) $, we have $ y(x_{1}^{i},x_{2}^{i}) =0$ and $ y(x_{1}^{i},x_{2}^{i})=1 $ for a negative and postive match respectively, where each training example is indexed by $ i $ and $ y $ represent the true label for an example. Our goal is to minimize the following loss function

\[
L_{s} = \frac{1}{m} \sum_{i=1}^{m} L(\hat{y}^{i},y^{i})
\]

where $ L(\hat{y},y) =  -y\log \hat{y} + (1-y)\log (1-\hat{y}) $ and $ \hat{y} = \hat{y}(x_{1},x_{2}) $ is the prediction of the network.

\subsection{Triplet network}
As shown in Figure~\ref{fig:architecture}, a triplet network consist of three replicas of a feedforward network that share identical weights. It accepts three input sentence, a positive, an anchor and a negative sentence, generates an encoding at layer P1 and then computes the distance between the anchor sentence and positive sentence, and anchor sentence and negative sentence.

\paragraph{Loss function}

Let $ x_{a}, x_{n} $ and $ x_{p} $ denote the anchor, negative and positive sentences respectively. Given a training instance, $ x_{a}, x_{n} $ and $ x_{p} $, we desire the distance between the encodings for $ x_a $ and $ x_p $ to be smaller than the distance between the encodings for $ x_a $ and $ x_n $ by a specified margin, m. In other words, for any $ x_p $ and $ x_n $, we aim to enforce 
\begin{equation}
\label{eqn:tripletconstraint}
d(x_a,x_p) + m < d(x_a,x_n)
\end{equation}
 
where $ d $ is some distance function. Thus, if we let g represent the encoding function, LM-P1, and use L2 norm as the distance function, the loss to be minimized is given as 

\[
 L_{t} = \sum_{i=1}^{m} \max (||g(x_a)-g(x_p)||^2 - ||g(x_a)-g(x_n)||^2 + m,0)
\]

where m connotes the mini-batch size.

\paragraph{Triplet selection}

To attain an optimal result, it is beneficial to consider mainly triplets that do not satisfy (\ref{eqn:tripletconstraint}) in every mini-batch. Using (\ref{eqn:tripletconstraint}), we can classify such triplets into three categories: Easy triplet, that already satisfy (\ref{eqn:tripletconstraint}); Hard triplets that satisfy $ d(x_a,x_n) < d(x_a,x_p) $ and Semi-hard triplets that satisfy $ d(x_a,x_p) < d(x_a,x_n) < d(x_a,x_p) + m $.

Some researchers~\cite{faghri2017} reported faster convergence when employing only negative triplets but this was 
contradicted in~\cite{schroff2015} where local minima issues was observed during the intial training phase for their problem set. The authors~\cite{schroff2015} proposed the use of semi-hard triplets to avoid such challenges. Now, each question in the WikiQA and TrecQA corpus is associated with a certain number of sentences that has being labelled as either a positive or negative match. In our case, the size of the training corpus is very moderate. Hence, considering all possible triplets with respect to a question and its corresponding candidate answer sentences was not expensive since the size of the newly generated corpus was a little over the original.


%

\section{Experiments}
\label{experiments}

\subsection{Datasets}

TrecQA \cite{wang2007} and WikiQA \cite{yang} datasets are commonly employed for the evaluation of answer selection models. All results presented in this work are based on this corpus.

TrecQA is a collection of question-answers pairs and the questions are factoid.This dataset was built from the Question Answering track (8-13) data of Text REtrieval Conference (TREC). For a given question, there exist a set of document from which candidate answer sentences are extracted from. A sentence is automatically selected based on pattern matching and the amount of non-stop words that overlap with the question. Finally, the relevance of each sentence with respect to a question is decided upon manually by annotators. TrecQA has a raw and clean version. The development and test sets of the clean version is a strict subset of the raw version's but the training sets of both variants are equal. In the clean version \cite{shen}, each question in the development and test set has at least one correct and one wrong answer in the corresponding set of candidate answer sentences. The clean version is used in this work.

The WikiQA \cite{yang} dataset  was generated from aggregated natural questions issued to the Bing search engine and sentences extracted from selected Wikipedia articles. Question-answer(s) pairs for which the question has at least one correct answer in the corresponding set of candidate answer sentence(s) is commonly used \cite{yang}. This setup is adopted in this work. The distribution of the number of question and candidate answer sentences in the training, development and test set for WikiQA and TrecQA are summarized in Table~\ref{tab:wikiqa_trecqa}.

\begin{table}[H]
	\centering
	\begin{threeparttable}
		\caption{WikiQA and TrecQA corpus statistics}
		\label{tab:wikiqa_trecqa}
		
		\begin{tabular}{lcccc}
			\toprule
			& \multicolumn{2}{c}{WikiQA} & \multicolumn{2}{c}{TrecQA} \\
			& \thead{Number of\\ questions}     & \thead{Number of\\ answers} & \thead{Number of\\ questions}     & \thead{Number of\\ answers} \\
			\midrule
			Train & $872$  & $8666$ & $89$  & $5914$   \\
			Dev     & $126$ & $1130$ & $69$ & $1343$     \\
			Test     & $241 $     & $2318$ & $68 $     & $1441$ \\
			\bottomrule
		\end{tabular}
		\begin{tablenotes}	
			\item 
		\end{tablenotes}
	\end{threeparttable}
\end{table}

\subsection{Evaluation metrics}

The performance of all models presented in this work was evaluated using the Mean Reciprocal Rank (MRR) and Mean Average Precision (MAP).

For a given number, $N$, of questions,
 supppose $R_{i}$ represents the position of the 
the first correct candidate answer
for question $q_{i}$. Then we obtain MRR as follows:
\[
 \text{MRR} = \frac{1}{N}\sum_{i=1}^N \frac{1}{R_{i}}
\]
The calculation of MAP is best understood by concentrating on how the Average precision (AP) is computed for a particular question, $q_{i}$. For this question, the precision score is computed for each correct answer sentences that appears in the search result and is obtained by dividing the number of correct answer sentences up to the current position by the total number of sentences up to the current position. The average across all correct answer sentence within the search results gives the AP. We can summarize this process in the definition below:
\[
\text{MAP} =  \frac{1}{N}\sum_{i=1}^N AP_{i}
\]

\subsection{Experimental setup}
ASBERT is built upon the PyTorch implementation of the following pretrained language models (LM): BERT, RoBERTa and XLNet. The maximum number of input tokens was set as $128$ and a batch size of $32$ was used throughout the experiments. For the triplet network architecture, the margin in the loss function was set as $5$. We employ an Adam optimizer with an initial learning rate of $2e-5$ and a linear warmup so that the learning rate linearly increases from $0$ to $2e-5$ after seeing $10$ percent of the data. The default epsilon value of $1e-8 $ was fixed. To prevent exploding gradient issues during training, the norm of the gradient values was clipped to $1$. Lastly, early stopping, with a patience of $20$ epochs, was adopted using the maximum MAP score for the Dev set.

\subsection{Results and discussion}
We report the performance of various state-of-the-art encoders and our encoding functions on the WikiQA and TrecQA dataset in Table \ref{tab:wikiqaresult} and \ref{tab:trecqaresult} respectively.

 GloVe~\cite{pennington2014glove} and fastText~\cite{bojanowski2016enriching} are models used to compute distributed word representations. To derive a sentence embedding, we average the vector representation of each word in the sentence. On the other hand, InferSent~\cite{conneau2017}, Universal Sentence Encoder (USE)~\cite{cer2018universal}, CLS-BERT, MEAN-BERT and Sentence-BERT~\cite{reimers2019} compute the vector representation of each text directly. CLS-BERT uses the BERT representation for the first token while MEAN-BERT employs the mean of the contextualized BERT embeddings for all tokens in the text. Siamese-BERT, Siamese-RoBERTa and Siamese-XLNet are variants of the architecture shown in Figure~\ref{fig:architecture}(a) where the LM module is BERT, RoBERTa and XLNet respectively. Similarly, Triplet-BERT, Triplet-RoBERTa and Triplet-XLNet are different versions of the network shown in Figure~\ref{fig:architecture}(b) where the LM module is BERT, RoBERTa and XLNet respectively. The model comparisons are based on the MAP of the dev set and test set, and the MRR of the test set.

For the WikiQA dataset, on average, the MEAN-BERT embeddings had the least performance while the USE embeddings was the most superior of the baseline models. Over all, Triplet-BERT had the best performance with a MAP score of $ 0.830 $ and $ 0.795 $ for the dev and test set respectively and an MRR score of 0.804 for the test set. These values are significantly greater than corresponding scores  of CLS-BERT and MEAN-BERT. We also observe that using the BERT pretrained model as the LM module in Figure~\ref{fig:architecture} produced the best results for both architectures. In addition, the triplet network not only achieved the best performance but pairwise comparison of both architectures initialized with the same pre-trained model demonstrates the superiority of the triplet network over the siamese except for the XLNet model.


\begin{table}[!htb]
\centering
\begin{threeparttable}
  \caption{Results on WikiQA dataset}
  \label{tab:wikiqaresult}
  \begin{tabular}{lccc}
    \toprule
Model & \multicolumn{2}{c}{MAP} & MRR \\
 & Dev & Test & Test \\
    \midrule
    GloVe & $ 0.571 $& $ 0.598 $& $ 0.611 $\\
    fastText & $ 0.612 $& $ 0.626 $& $ 0.638 $\\
    InferSent & $ 0.447 $& $ 0.474 $& $ 0.479 $\\
    USE & $ \bm{0.614} $& $\bm{ 0.648} $& $ \bm{0.655} $\\
    CLS-BERT & $ 0.447 $& $ 0.474 $& $ 0.479 $ \\
    MEAN-BERT & $ 0.343 $& $ 0.337 $& $ 0.341 $\\
    Sentence-BERT & $ 0.564 $& $ 0.586 $& $ 0.593 $\\
    \midrule
    Siamese-BERT & $0.767$ & $0.704$ &  $0.719$\\
    Siamese-RoBERTa &$0.739$ & $0.685$ &  $0.697$ \\
    Siamese-XLNet &$0.736$ & $0.683$ &  $0.697$ \\
    Triplet-BERT & $\bm{0.830}$ & $\bm{0.795}$ &  $\bm{0.804}$ \\
    Triplet-RoBERTa & $0.817$ & $0.787$ &  $0.803$ \\
    Triplet-XLNet & $0.724$ & $0.665$ &  $0.677$ \\
    \bottomrule
  \end{tabular}
  \begin{tablenotes}
  
   \item 
    \end{tablenotes}
    \end{threeparttable}
\end{table}

We also compared the performance of the baseline models with ours on the TrecQA dataset and the result is shown in Table \ref{tab:trecqaresult}. The same settings for all models used on WikiQA was also applied for TrecQA datasets. Analogous to the performance on WikiQA datasets, the USE embeddings achieved the best result over all baseline models while MEAN-BERT was the least.The best mean result amongst our models was realized by Triplet-BERT similar to its performance on the WikiQA dataset. We also observe embeddings from the triplet networks appear to be more superior than their corresponding siamese counterparts. The only oddity seen with an XLNet LM module where siamese network produced better embeddings than the triplet network. In addition, we again observe that for both siamese and triplet network, the BERT LM module outperformed other pretrained models.

\begin{table}[!htb]
\centering
\begin{threeparttable}
  \caption{Results on TrecQA dataset}
  \label{tab:trecqaresult}
  \begin{tabular}{lccc}
    \toprule
Model & \multicolumn{2}{c}{MAP} & MRR \\
 & Dev & Test & Test \\
    \midrule
    GloVe & $ 0.579 $& $ 0.511 $& $ 0.602 $\\
    fastText & $ 0.629 $& $ 0.559 $& $ 0.632 $\\
    InferSent & $ 0.534 $& $ 0.585 $& $ 0.671 $\\
    USE & $ \bm{0.716} $& $ \bm{0.707} $& $ \bm{0.819} $\\
    CLS-BERT & $ 0.534 $& $ 0.585 $& $ 0.671 $\\
    MEAN-BERT & $ 0.391 $& $ 0.363 $& $ 0.429 $\\
    Sentence-BERT & $ 0.580 $&$  0.589 $& $ 0.691 $\\
    \midrule
    Siamese-BERT & $0.627$ & $0.604$ &  $0.650$\\
    Siamese-RoBERTa & $ 0.634 $  & $ 0.583 $  & $ 0.635 $   \\
    Siamese-XLNet &$0.620$ & $0.527$ &  $0.559$ \\
    Triplet-BERT &  $ \bm{0.674} $&  $ \bm{0.752} $&  $ \bm{0.835} $ \\
    Triplet-RoBERTa & $0.687$ & $0.736$ &  $0.827$ \\
    Triplet-XLNet & $0.551$ & $0.567$ &  $0.645$ \\
    \bottomrule
  \end{tabular}
  \begin{tablenotes}
  
   \item 
    \end{tablenotes}
    \end{threeparttable}
\end{table}

To gain an intuition on the quality of the encodings, we observe the top three results of two randomly chosen questions from the WikiQA test set. One of the questions is factoid and the other is non-factoid, and candidate sentences were pulled from the same corpus. Table \ref{tab:qualitative} shows the search results using embeddings derived from the best settings of both siamese (MAP = $ 0.704 $) and triplet (MAP = $ 0.795 $) architecture. In general, it appears the triplet network produces better embeddings since it has more relevant candidate answer sentences ranked higher.

\begingroup
\renewcommand{\arraystretch}{1.5}
\begin{table}[H]
	\centering
	\begin{threeparttable}
		\caption{Search results for questions chosen at random from the WikiQA test set }
		\label{tab:qualitative}
		{\scriptsize
			\begin{tabular}{lp{3.6cm}p{4.6cm}}
				\toprule
				Question & Siamese ($\text{MAP} =  0.704 $)    & Triplet ($\text{MAP} = 0.795 $)   \\
				\midrule
				who is the founder of twitter & 
				
				Twitter is an online social networking service and microblogging service that enables its users to send and read text-based messages of up to 140 characters , known as "tweets" 
				& Twitter was created in March 2006 by Jack Dorsey and by July, the social networking site was launched     \\
				&Twitter was created in March 2006 by Jack Dorsey and by July, the social networking site was launched &Twitter is an online social networking service and microblogging service that enables its users to send and read text-based messages of up to 140 characters , known as "tweets"\\
				&Twitter Inc. is based in San Francisco , with additional servers and offices in New York City , Boston , and San Antonio &Twitter Inc. is based in San Francisco , with additional servers and offices in New York City , Boston , and San Antonio\\
				
				how does interlibrary loan work & 
				
				The end user is usually responsible for any fees, such as costs for postage or photocopying 
				& Interlibrary loan (abbreviated ILL, and sometimes called interloan, document delivery, or document supply) is a service whereby a user of one library can borrow books or receive photocopies of documents that are owned by another library     \\
				&The lending library usually sets the due date and overdue fees of the material borrowed & The term document delivery may also be used for a related service, namely the supply of journal articles and other copies on a personalized basis, whether these come from other libraries or direct from publishers\\
				&Commercial document delivery services will borrow on behalf of any customer willing to pay their rates& The user makes a request with their local library, which, acting as an intermediary, identifies owners of the desired item, places the request, receives the item, makes it available to the user, and arranges for its return\\
				
				\bottomrule
			\end{tabular}
		}
		\begin{tablenotes}	
			\item 
		\end{tablenotes}
	\end{threeparttable}
\end{table}
\endgroup

\section{Conclusion}
\label{conclusion}

This paper presented a framework to learn meaningful text encodings to address answer selection (AS) task. The framework consists of two architectures - a siamese and triplet network. The siamese network treats AS problem as a supervised learning task where the input is a question and a candidate answer sentence and output is a label which connotes a positive or negative match. The triplet network, on the other hand, accepts three input texts - an anchor, a positive and a negative sentence. It then applies a distance function to an encoding of each sentence, towards the end of the network, to determine their closeness. Once trained optimally, the piece of the network used to generate a vector repesentation is retained.

The performance of our approach was compared with other state-of-the-art encoding functions using the TrecQA and WikiQA datasets. The experimental results demonstrates that our method outperforms all strong baselines considered.

%

\medskip
\nocite{*}

\end{document}